\documentclass[11pt,a4paper]{article}
\usepackage[utf8]{inputenc}
\usepackage[T1]{fontenc}
\usepackage{lmodern}
\usepackage{amsmath,amssymb,amsthm}
\usepackage{geometry}
\usepackage{hyperref}
\usepackage{booktabs}
\usepackage{microtype}
\usepackage{cite}

\hypersetup{
    colorlinks=true,
    linkcolor=blue,
    citecolor=blue,
    urlcolor=blue,
    pdfauthor={Volkan Dagli, Zerrin Dagli, Daghan Dagli},
    pdftitle={Bipartite Carrier Compression and Machine-Certified Bounds in Schonhage Trilinear Multiplication Tensors}
}

\newtheorem{theorem}{Theorem}

\title{\textbf{Bipartite Carrier Compression and Machine-Certified Bounds in Sch\"{o}nhage Trilinear Multiplication Tensors}}

\author{
    \textbf{Volkan Da\u{g}l\i}$^{1,2}$ \quad 
    \textbf{Zerrin Da\u{g}l\i}$^{3,4}$ \quad 
    \textbf{Da\u{g}han Da\u{g}l\i}$^{5}$\\[0.5em]
    \small $^{1}$Anadolu University, Eski\c{s}ehir, T\"{u}rkiye \\
    \small $^{2}$ITouch Systems, Adana, T\"{u}rkiye \\
    \small $^{3}$Mersin University, Mersin, T\"{u}rkiye \\
    \small $^{4}$Yusuf Kalkavan Anadolu Lisesi, Mersin, T\"{u}rkiye \\
    \small $^{5}$Toros Science High School, MEV (Toros University), Mersin, T\"{u}rkiye\\[0.4em]
    \small \texttt{ORCID: 0009-0000-1587-8703 (V.D.), 0000-0001-9490-6425 (Z.D.), 0009-0003-2492-8313 (D.D.)}
}

\date{\today}

\begin{document}

\maketitle

\begin{abstract}
Under the conditional framework breaking the longstanding $O(n \log n)$ integer multiplication barrier (OpenAI Preprint \#109), the asymptotic time complexity exponent $T(n) = O(n (\log n)^{1-\kappa})$ is governed by bounding the second moment $\mathbb{E}[X^2] < 1$ of recursive branching processes over trilinear tensor carrier graphs. In this work, we present a constructive combinatorial optimization over the alternating carrier bipartite matching graphs of the Sch\"{o}nhage trilinear tensor decomposition at dimensions $h=23$ and $h=25$. Through an alternating augmenting path search, we uncover a non-trivial cluster-unlocking cascade that eliminates 185 carrier roles (95 roles in $h=23$ and 90 roles in $h=25$), reducing the global physical wire width by 377,890 wires from the reviewed baseline $W = 177,284,805$ down to $W = 176,906,915$. Evaluated over the exact $10^{-14}$ rational grid, this wire reduction certifies an improved second-moment coefficient $AB = \frac{5,164,059}{125,000,000,000}$ and establishes the sharpened conditional exponent
\[
\kappa = \frac{826,215,307}{20,000,000,000,000} \approx 4.131076535 \times 10^{-5},
\]
strictly surpassing the preceding record on this architecture ($\kappa \approx 4.123864 \times 10^{-5}$). All 47 polynomial constraints and 7 rational security margins are audited in exact $\mathbb{Q}$-arithmetic with zero floating-point error, and the fundamental wire reduction identity is formally verified in the Lean 4 interactive theorem prover without unproven axioms.
\end{abstract}

\section{Introduction}

The computational complexity of multiplying two $n$-bit integers has been a central problem in theoretical computer science since Karatsuba's 1960 discovery of subquadratic multiplication \cite{karatsuba1962}. After decades of refinement through the Sch\"{o}nhage--Strassen algorithm \cite{schonhage1971} and F\"{u}rer's bound \cite{furer2009}, Harvey and Hoeven established the celebrated $O(n \log n)$ conditional upper bound \cite{harvey2019}. 

Recently, OpenAI Preprint \#109 proposed an analytical machinery breaking the $O(n \log n)$ barrier, demonstrating that integer multiplication admits an upper bound of the form:
\begin{equation}
T(n) = O\left(n (\log n)^{1 - \kappa}\right), \quad \kappa > 0.
\end{equation}
In this framework, the exponent $\kappa$ is constrained by establishing a finite arithmetic certificate ensuring that a certain second moment $\mathbb{E}[X^2]$ remains strictly below unity across recursive branching levels. The value of $\kappa$ depends inversely on the global physical wire width $W$, which measures the total carrier cost of operations across the tensor product decomposition.

Community-driven research coordinated around the \texttt{CrocSwap/integer-mult-bounds} repository \cite{colkitt2026} established successive improvements to the carrier graph. In particular, the branch documented in PR \#48 and PR \#49 (Arun \cite{arun2026}) achieved $\kappa = \frac{4,123,863,984}{10^{14}} \approx 4.123864 \times 10^{-5}$ by searching for carrier reductions in the trilinear tensor components at $h=23$ and $h=25$, arriving at a wire width $W = 177,284,805$.

In this paper, we demonstrate that the carrier assignment graph in the Sch\"{o}nhage alternating tensor decomposition possesses latent augmenting paths that were previously unreached. By formulating a multi-round greedy climber with alternating parity probing, we achieve a substantial reduction in both $h=23$ and $h=25$, yielding an exact, machine-verified record for this tensor class.

\section{Sch\"{o}nhage Alternating Carrier Tensor Architecture}

\subsection{Wire Width Formulation}

In the trilinear tensor framework parameterized by coprime dimensions $a=23$ and $b=25$, the base modulus is $m = a \cdot b = 575$. The base combinatorial size is $N = \binom{a}{3}\binom{b}{3} = \binom{23}{3}\binom{25}{3} = 1,771 \times 2,300 = 4,073,300$. The global physical wire width $W$ is affine in the active carrier role counts $R_{23}$ and $R_{25}$:
\begin{equation}
W(R_{23}, R_{25}) = 2N + \frac{N}{\binom{23}{3}} R_{23} + \frac{N}{\binom{25}{3}} R_{25} = 8,146,600 + 2,300 \cdot R_{23} + 1,771 \cdot R_{25}.
\end{equation}
Here, the fixed blocks $2N = 8,146,600$ originate from the boundary and center data routings. The integer multipliers:
\begin{itemize}
    \item $w_{23} = 2,300$ per role in dimension $h=23$,
    \item $w_{25} = 1,771$ per role in dimension $h=25$,
\end{itemize}
dictate that role eliminations in $h=23$ yield higher wire savings per role than in $h=25$.

Prior to our work, the minimal verified carrier counts reported by Arun \cite{arun2026} were $R_{23} = 36,382$ and $R_{25} = 48,255$, which yielded:
\[
W_{\text{PR49}} = 8,146,600 + 2,300 \cdot (36,382) + 1,771 \cdot (48,255) = 177,284,805.
\]

\subsection{Bipartite Carrier Matching}
Each component $h \in \{23, 25\}$ is evaluated as an alternating carrier graph $G_h = (V, E)$, wherein a subset of carrier roles is matched to cancel cross-terms in the trilinear tensor decomposition. Maximizing the size of the matching directly minimizes the uncancelled carrier role count $R_h = |V| - |\mathcal{M}_h|$.

\section{Augmenting Carrier Optimization and the Cascading Transition}

\subsection{Optimization Protocol}
We designed a 15-round augmenting carrier search procedure operating on the bipartite matching state. In each round:
\begin{enumerate}
    \item Candidate swap moves are evaluated across tensor parity bits (\texttt{tbits}) and vertex permutation bits (\texttt{vbits}).
    \item Maximum cardinality matchings are re-computed via Hopcroft--Karp augmentation with alternating parity orientation checks.
    \item Moves that strictly decrease $R_h$ while satisfying CRT consistency ($p = 2^{61}-1$, zero disagreements) and rank mass conservation ($\sum r \cdot n_r = h \cdot R + 2 \cdot \text{loss}$) are accepted.
\end{enumerate}

\begin{table}[ht]
\centering
\caption{Detailed 15-Round Optimization Trajectory for Dimension $h=25$.}
\label{tab:rounds}
\vspace{0.5em}
\begin{tabular}{cccccc}
\toprule
\textbf{Round} & \textbf{Winning Move} & \textbf{New $R_{25}$} & \textbf{Role Drop} & \textbf{Matched Roles} & \textbf{Execution (s)} \\
\midrule
1 & \texttt{tbit 28} & 48,247 & $-8$ & 7,817 & 464.2 \\
2 & \texttt{vbit 77} & 48,239 & $-8$ & 7,833 & 465.6 \\
3 & \texttt{vbit 88} & 48,231 & $-8$ & 7,849 & 468.1 \\
4 & \texttt{tbit 55} & 48,225 & $-6$ & 7,855 & 467.4 \\
5 & \texttt{vbit 99} & 48,219 & $-6$ & 7,867 & 473.3 \\
6 & \texttt{vbit 110} & 48,213 & $-6$ & 7,879 & 469.9 \\
7 & \texttt{vbit 121} & 48,209 & $-4$ & 7,887 & 471.0 \\
8 & \texttt{vbit 154} & 48,205 & $-4$ & 7,895 & 472.5 \\
9 & \texttt{tbit 100} & 48,203 & $-2$ & 7,897 & 474.1 \\
10 & \texttt{tbit 166} & 48,201 & $-2$ & 7,899 & 474.8 \\
11 & \texttt{tbit 196} & 48,196 & $-5$ & 7,904 & 476.3 \\
12 & \texttt{tbit 187} & 48,188 & $-8$ & 7,912 & 477.0 \\
13 & \texttt{tbit 178} & 48,179 & $-9$ & 7,921 & 478.4 \\
14 & \texttt{tbit 193} & 48,167 & $-12$ & 7,933 & 479.2 \\
15 & \texttt{vbit 143} & \textbf{48,165} & $-2$ & \textbf{7,937} & 481.5 \\
\bottomrule
\end{tabular}
\end{table}

\subsection{The Cluster Unlocking Phenomenon}
A notable structural phenomenon occurred during the $h=25$ trajectory (Table~\ref{tab:rounds}). In rounds 7--10, the gain decayed smoothly from $-4$ to $-2$, suggesting an approaching local optimum. However, at round 11, the flip of \texttt{tbit 196} altered an alternating parity condition, unblocking a large family of augmenting paths. This triggered an accelerating cascade in rounds 12--14 ($-8 \to -9 \to -12$ roles), demonstrating that carrier graph topologies exhibit non-convex multi-valley structures.

\subsection{Net Carrier and Wire Reductions}
Across all rounds, the results achieved are:
\begin{itemize}
    \item \textbf{Dimension $h=23$:} $R_{23}$ reduced from $36,382$ to $\mathbf{36,287}$ ($-95$ roles, $+143$ matched).
    \item \textbf{Dimension $h=25$:} $R_{25}$ reduced from $48,255$ to $\mathbf{48,165}$ ($-90$ roles, $+128$ matched).
    \item \textbf{Cumulative Roles Saved:} $95 + 90 = \mathbf{185}$ carrier roles eliminated.
    \item \textbf{Total Wire Savings:}
    \[
    \Delta W = 95 \times 2,300 + 90 \times 1,771 = 218,500 + 159,390 = \mathbf{377,890\text{ wires}}.
    \]
    \item \textbf{Certified Wire Width:}
    \[
    W_{\text{certified}} = 177,284,805 - 377,890 = \mathbf{176,906,915}.
    \]
\end{itemize}

\section{Exact Arithmetic Verification and Certified Bounds}

\subsection{\texorpdfstring{Second-Moment Parameter $AB$ Bisection}{Second-Moment Parameter AB Bisection}}
With $W = 176,906,915$, the second-moment condition $\text{moment}(m, W, \text{child\_multiplicities}, AB)['\text{upper}'] < 1$ was evaluated via bisection over the rational grid $\mathbb{Q} \cap 10^{-14} \mathbb{Z}$. The maximal certified rational numerator is:
\begin{equation}
AB_{\text{certified}} = \frac{4,131,247,200}{10^{14}} = \frac{5,164,059}{125,000,000,000} \approx 4.1312472 \times 10^{-5}.
\end{equation}

\subsection{\texorpdfstring{Certified Exponent $\kappa$ Formulation}{Certified Exponent kappa Formulation}}
Through assembly of the finite bridge representation and evaluation of the 47 polynomial constraints governing the branching process, we obtain the certified exponent:
\begin{equation}
\kappa_{\text{certified}} = \frac{826,215,307}{20,000,000,000,000} \approx \mathbf{4.131076535 \times 10^{-5}}.
\end{equation}

The verification was completed through the following criteria:
\begin{enumerate}
    \item \textbf{47/47 Strict Inequalities:} Every constraint in the algebraic assembly evaluates to a strictly positive rational number ($> 0$).
    \item \textbf{7/7 Security Margins:} All seven slack margins $\{g_1, \dots, g_7\}$ remain strictly positive, guaranteeing robustness of the numerical witness.
    \item \textbf{Exclusion of Preceding Configurations:} All prior configurations (PR \#40 through PR \#48) are strictly excluded at this threshold (lower moments $> 1$).
    \item \textbf{Grid Boundary Rejection:} The subsequent grid point $\kappa + 10^{-14}$ evaluates to an upper bound $\ge 1$, formally verifying that $\kappa_{\text{certified}}$ is the optimal grid solution.
\end{enumerate}

\section{Lean 4 Formal Verification}

To ensure complete mathematical integrity and eliminate reliance on floating-point arithmetic or runtime interpreter bugs, we formalized the tensor wire width equations and the strict superiority theorem in the Lean 4 interactive theorem prover (Mathlib v4.11.0 compatible).

\begin{theorem}[Lean 4 Machine-Checked Wire Reduction]
Let $W(r_{23}, r_{25}) = 2 \cdot 4073300 + 2300 \cdot r_{23} + 1771 \cdot r_{25}$. Then:
\[
W(36287, 48165) + 377890 = W(36382, 48255),
\]
and $W(36287, 48165) = 176,906,915$.
\end{theorem}

\begin{theorem}[Lean 4 Strict Improvement Theorem]
Let $\kappa_{\text{certified}} = \frac{826,215,307}{20,000,000,000,000}$, $\kappa_{\text{PR49}} = \frac{4,123,863,984}{10^{14}}$, and $\kappa_{\text{PR48}} = \frac{4,118,625,410}{10^{14}}$. Then:
\[
\kappa_{\text{certified}} > \kappa_{\text{PR49}},
\]
and furthermore:
\[
(\kappa_{\text{certified}} - \kappa_{\text{PR49}}) > (\kappa_{\text{PR49}} - \kappa_{\text{PR48}}).
\]
\end{theorem}

Both theorems are fully proven in \texttt{SchonhageTensorMatch.lean} with zero reliance on unproven assumptions (\texttt{sorry}), verified deterministically via \texttt{norm\_num}.

\section{Conclusion and Open Perspectives}

We have established a new record for the Sch\"{o}nhage alternating carrier tensor track in integer multiplication complexity, eliminating 185 carrier roles, saving 377,890 physical wires, and improving the conditional exponent to $\kappa \approx 4.1310765 \times 10^{-5}$. The observation of the cluster unlocking phenomenon indicates that alternating carrier graphs possess deeper topological structures that warrant further exploration. Complete artifacts, witness verifiers, and Lean 4 proofs are deposited openly under reproducible open-access licenses \cite{zenodo2026}.

\section*{Acknowledgments}
The authors express their sincere gratitude to the theoretical computer science and computational complexity research community, particularly the maintainers and contributors of the \texttt{integer-mult-bounds} collaborative platform.

We gratefully acknowledge Bekirhan Da\u{g}l\i{} (Managing Partner, ITouch Systems) and ITouch Systems Research \& Development for providing the dedicated bare-metal server infrastructure, high-throughput virtualization clusters, and sustained computational resources essential for executing the multi-round alternating carrier search and formal verification pipelines.

\end{document}